\documentclass[%
 aip,
 amsmath,amssymb,
 reprint,%
]{revtex4-2}

\usepackage{graphicx}
\usepackage{dcolumn}
\usepackage{bm}
\usepackage{physics}
\usepackage{braket}
\usepackage[colorlinks=true]{hyperref}

\usepackage[utf8]{inputenc}
\usepackage[T1]{fontenc}
\usepackage{mathptmx}

\begin{document}

\preprint{AIP/123-QED}

\title[Superconducting electro-mechanics to test Di\'osi-Penrose effects of general relativity in massive superpositions]{Superconducting electro-mechanics to test Di\'osi-Penrose\\ effects of general relativity in massive superpositions}
\author{Mario F. Gely}
\email{mario.gely@physics.ox.ac.uk}
\affiliation{%
Kavli Institute of NanoScience, Delft University of Technology,\\
PO Box 5046, 2600 GA, Delft, The Netherlands.
}
\author{Gary A. Steele}

\affiliation{%
Kavli Institute of NanoScience, Delft University of Technology,\\
PO Box 5046, 2600 GA, Delft, The Netherlands.
}%

\date{\today}

\begin{abstract}
Attempting to reconcile general relativity with quantum mechanics is one of the great undertakings of contemporary physics.
Here we present how the incompatibility between the two theories arises in the simple thought experiment of preparing a heavy object in a quantum superposition.
Following Penrose's analysis of the problem, we determine the requirements on physical parameters to perform experiments where both theories potentially interplay.
We use these requirements to compare different systems, focusing on mechanical oscillators which can be coupled to superconducting circuits.
\end{abstract}

\maketitle
One problem in combining the theories of general relativity (GR) and quantum mechanics (QM) concerns the passage of time.
In GR, the rate at which time passes can be distorted by the presence of a massive object, a phenomenon called time-dilation.
In QM, an object can exist in a superposition of being at two locations at the same time.
Together, these effects could give rise to a space-time superposition in which time flows at two different rates.
This situation cannot be described using QM, where time is not a quantum operator.
For this reason, performing such an experiment could lead to the discovery of new physics shedding light on the details of a theory unifying GR and QM.
On the flip-side, no experimentally proven theory of quantum gravity is available to predict which physical parameters may lead to GR-induced deviations from QM, and the multiple approaches to the problem remain theoretical~\cite{Gorelik_2005,feynman1962,karolyhazy1966gravitation,DIOSI1987377,ellis1989quantum,ghirardi1990continuous,chaves1994metric,penrose1996gravity,PhysRevD.54.1600,power2000decoherence,milburn2006lorentz,van2008schrodinger,Anastopoulos_2013,PhysRevLett.111.021302,oosterkamp2013clock,diosi2005intrinsic,PhysRevD.93.044027,PhysRevD.93.044027,BRUSCHI2016182}.
One prominent theory was put forth by Penrose~\cite{penrose1996gravity}, hypothesizing that superpositions experience a collapse of their wave-function due to GR. 
Penrose provided a time-scale associated with this effect, identical (up to a numerical factor) to that introduced by Di\'osi~\cite{DIOSI1987377} though based on different arguments.
The so-called Di\'osi-Penrose model is then one instance of the broader class of collapse models~\cite{bassi2013models}, and would give a fundamental mass-related origin to the quantum-to-classical transition, addressing foundational questions in the interpretation of quantum mechanics.
Exploring the interplay of general relativity and quantum mechanics could thus be achieved by studying massive quantum superpositions.
Two criteria can help in estimating if an experimental system is suitable
(i) techniques should be available to prepare a large spatial superposition of the system, and monitor its evolution, 
(ii) the estimated time-scale on which general-relativistic effects come into play should be smaller than the coherence time of the quantum state.
Concerning criteria (i), clamped mechanical oscillators coupled to superconducting circuits show promise~\cite{schwab2005putting}.
Through optomechanical coupling of motion to a superconducting resonator~\cite{aspelmeyer2014cavity}, sideband cooling can force the mechanical oscillator into its ground-state~\cite{Teufel2011}.
For mechanical oscillators with large zero-point fluctuations, the evolution of the ground-state may already be influenced by general-relativistic effects.
But superconducting circuits can also introduce higher levels of non-linearity, enhancing control over quantum states of motion and the measurement of their evolution.
This could improve the prospects of detecting unconventional decoherence mechanisms~\cite{abdi2016dissipative}, hypothetically induced by general-relativistic effects, or prepare spatial superpositions exceeding the zero-point fluctuations of motion.
Using optomechanics, or photon-pressure coupling, entangled~\cite{ockeloen2018stabilized} and squeezed states of motion~\cite{wollman2015quantum,pirkkalainen2015squeezing} have been demonstrated, and coupling through superconducting interference shows promise~\cite{rodrigues2019coupling,kounalakis2019synthesizing,kounalakis2020flux}.
Alternatively, Josephson junctions can introduce sufficient levels of non-linearity to access the quantum states of mechanical harmonic oscillators.
This has been achieved by coupling acoustic resonators to qubits through piezoelectricity~\cite{o2010quantum,satzinger2018quantum,chu2018creation,arrangoiz2019resolving} or through the current created by the movement of a charged oscillator~\cite{ma2020nonclassical}.
Finally, superconducting qubits can also play the role of remote quantum-state sources, with optomechanical interaction performing electronic to mechanical conversion of quantum states~\cite{reed2017faithful}.
Due to the strong experimental activity in creating quantum states of motion using superconducting circuits, we focus here on criteria (ii), and evaluate different mechanical oscillators on their potential for exploring the interaction between quantum mechanics and general relativity.

This article is divided into three parts.
First, we present how quantum superpositions of a massive object constitute a problem for our current theory of quantum mechanics when time-dilation is taken into account.
We will use a pedagogical gedanken experiment to introduce the time-scale on which Penrose predicted effects of general relativity to arise~\cite{penrose1996gravity}.
Secondly, we apply these concepts to mechanical oscillators.
We will write the different conditions that these oscillators should obey in order to test the previously laid out hypotheses.
Thirdly, we will use these conditions to compare typical mechanical oscillators that can be coupled to superconducting circuits: acoustic resonators, beams and membranes.

\section{GR effects in massive quantum superpositions}
Following pioneering work on quantum gravity in the early 20th century by Matvei Bronstein~\cite{Gorelik_2005}, the possible impact of general relativity on quantum mechanical coherence was then put forth by Feynman~\cite{feynman1962}, and many authors have since elaborated on this idea \cite{karolyhazy1966gravitation,DIOSI1987377,ellis1989quantum,ghirardi1990continuous,chaves1994metric,penrose1996gravity,power2000decoherence,milburn2006lorentz,van2008schrodinger,oosterkamp2013clock,BRUSCHI2016182,bruschi2020self}.
Some of these works, as well as methods to test them, are reviewed in detail in Refs.~\cite{diosi2005intrinsic,bassi2013models,carney2019tabletop}.
Here, we focus on the theory of Penrose~\cite{penrose1996gravity}.
To give a pedagogical introduction to the topic, we first present a heuristic derivation of the time-scale on which general-relativity may influence a quantum superposition, consistent with the theory of Penrose, by simplifying the thought experiment presented in Ref.~\cite{oosterkamp2013clock}.

We consider an incompressible ball with mass $m$ and radius $R$.
Solving Einsteins field equations~\cite{schwarzschild1999gravitational} reveals that the proper time $t$ measured by a stationary clock at a distance $r$ from the center of the ball is related to the time $t_\text{far}$ measured by a clock infinitely far from the ball through
\begin{equation}
    t = t_\text{far}\sqrt{1-\frac{2mG}{rc^2}}\simeq t_\text{far}\left(1-\frac{mG}{rc^2}\right)\ ,
\end{equation}
where $G$ is the gravitational constant, $c$ is the speed of light and $r>R$.
The approximation holds for the small masses we will consider here $m\ll Rc^2/G$.

\begin{figure}[t]
\centering
\includegraphics[width=0.45\textwidth]{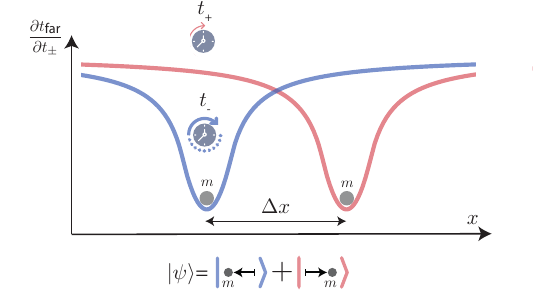}
\caption{\textbf{Uncertainty in time of a massive superposition.}
We consider a ball of mass $m$ in a quantum superposition $\ket{\psi}$ of being displaced in two opposite directions.
The two parts of the superposition are given a different color, blue and red.
We plot the rate at which the proper time $t_\pm$ corresponding the left (right) displaced ball evolves with respect to a the proper time $t_\text{far}$ of a reference location far away from the ball.
In the vicinity of one of the parts of the superposition, there is thus an ambiguity in the local time one should consider for the quantum evolution of the state quantified by the difference between $t_-$ and $t_+$.
}
\label{fig:time_dilation}
\end{figure}

The incompatibility with QM comes if we would place this ball in a superposition.
Specifically, if the ball is simultaneously displaced left and right as schematically shown in Fig.~\ref{fig:time_dilation}.
There is then no unique definition of time\footnote{Note that progress is being made on this topic, for example using quantum reference frames formalism~\cite{giacomini2021spacetime}} -- which is required to describe this system with Schr\"odinger's equation.
For example, there is an ambiguity in the time $t$ that a clock positioned near the left-displaced part of the superposition would record: does it experience the proper-time defined by the left- or right- displaced ball?
If it experiences the left-displaced time, then $t=t_- \simeq t_\text{far}(1-mG/Rc^2)$, if it experiences the right-displaced time, then  $t=t_+ \simeq t_\text{far}$ (if the spatial separation between the two parts of the superposition $\Delta x \gtrsim R$). 
The fact that there is an issue in describing this situation with our current physical theories is undeniable. 
What is currently debatable is for what system parameters the consequences of this situation start to become measurable.

We now estimate the parameters $m,R$ necessary for time-dilation to potentially have a noticeable effect in this superposition.
We label the states of the ball displaced to the left and right $\psi_\pm$, which we assume are eigenstates of a Hamiltonian describing this situation, with identical eigenenergies $E_\pm = E$.
According to Schr\"odinger's equation, the state of the system would then evolve following 
\begin{equation}
    \psi(t) = \left(\psi_+e^{itE/\hbar}+\psi_-e^{itE/\hbar}\right)/\sqrt{2}\ .
\end{equation}
However, if we take into account time-dilation, the time $t$ no longer has a clear definition.
We can only estimate the uncertainty in time $\Delta t = t_+-t_- $, which simplifies in the case $\Delta x \gtrsim 2R$ to
\begin{equation}
    \Delta t \simeq t_\text{far}Gm/Rc^2\ .
\end{equation}
This could result in an uncertainty in the relative phase of the two parts of this superposition $\Delta\theta = \Delta tE/\hbar$.
The time-scale on which this phase uncertainty may become significant (i.e. $\Delta\theta\sim2\pi$), we will call the general-relativistic, or GR time-scale $t_\text{GR}$.
Most of our ignorance in determining $t_\text{GR}$ lies in choosing of a value for $E$.
This absolute energy of the state would usually be neglected as a global phase which plays no role in the physics of the system.
Here it sets the time-scale on which GR effects come into play.
As put forth in Ref.~\cite{oosterkamp2013clock}, if one chooses this energy to be the relativistic rest-energy of the mass $E = mc^2$, one obtains a very similar time-scale to that obtained through different means~\cite{diosi1987favor,penrose1996gravity}.
In this assumption, we arrive at a time-scale for GR effects
\begin{equation}
    t_\text{GR,H} = h\frac{R}{Gm^2}\ ,
    \label{eq:t_G_heuristic}
\end{equation}
defined by $\Delta\theta(t_\text{far}=t_\text{GR,H}) = 2\pi$.
The subscript H refers to the heuristic nature of this derivation.
A requirement to observe deviations from QM in this time-scale is to have $t_\text{GR,H}\ll t_\text{coh}$, with $t_\text{coh}$ the coherence time of the system as determined by its interaction with the environment following conventional decoherence mechanisms~\cite{zurek2003decoherence}.
Note that the time $t_\text{GR,H}$ is crucially dependent on the guess $E = mc^2$ which is a very large number.
Whilst only experiments can validate or invalidate this guess, it is worth noting that other approaches to this problem have led to a similar estimates of the time-scale.
What will occur if a quantum superposition is sustained for this time-scale is debatable.
The fact that an uncertainty in phase could emerge from time-dilation points however towards decoherence. 
Whilst the heuristic arguments presented above are inspired by Ref.~\cite{oosterkamp2013clock}, a similar time-scale can be derived following Ref.~\cite{penrose1996gravity}.
In Ref.~\cite{penrose1996gravity}, Penrose quantifies the difference of free-falls through the space-times of two different parts of a superposition to justify a GR time-scale
\begin{equation}
\begin{split}
    t_\text{GR,P} = &\frac{\hbar}{\Delta E}\ ,
    \label{eq:t_G_penrose_full}
\end{split}
\end{equation}
related to the gravitational self-energy difference $\Delta E$
\begin{equation}
\begin{split}
    \Delta E =& 2E_{1,2}-E_1-E_2\ ,\\
    E_{i,j}=&4\pi G\int\int d\vec{r}_1 d\vec{r}_2 \frac{\rho_i(\vec{r}_1)\rho_j(\vec{r}_2)}{|\vec{r}_1-\vec{r}_2|}\ ,
    \label{eq:delta_E}
\end{split}
\end{equation}
where $\rho_{1,2}$ correspond to the mass distribution of the different parts of a superposition.
By considering the case of a ball of mass $m$, radius $R$, and uniform density, with a center of a mass in a spatial superposition over a distance $\Delta x\ge2R$~\cite{kleckner2008creating}
\begin{equation}
    \Delta E = 8\pi Gm^2\left(\frac{6}{5R}-\frac{1}{\Delta x}\right)\ .
    \label{eq:delta_E_calculated}
\end{equation}
We notice then that for $\Delta x\gg2R$, the separation in distance $\Delta x$ has little influence and we can write the corresponding time-scale as
\begin{equation}
    t_\text{GR,P} \simeq \hbar\frac{5R}{48\pi Gm^2}\ ,
    \label{eq:t_GRP}
\end{equation}
and distances $\Delta x\sim 2R$ would bring corrections on the order unity to this time-scale.
This result is very similar to that obtained heuristically in Eq.~(\ref{eq:t_G_heuristic}).

Other theoretical approaches have led to similar time-scales.
Most notably Di{\'o}si and Luk{\'a}cs~\cite{diosi1987favor} estimated the quantum fluctuations of a gravitational field in a given volume from the quantum-induced measurement imprecision of an accelerating ball in the field.
They then considered that these fluctuations in the gravitational field act as a noise source acting on massive objects which will cause decoherence~\cite{DIOSI1987377}.
The resulting decoherence time-scale is very similar to the one derived by Penrose~\cite{penrose1996gravity,bassi2013models}

\section{Implementation with mechanical oscillators}
\begin{figure*}[t]
\centering
\includegraphics[width=0.9\textwidth]{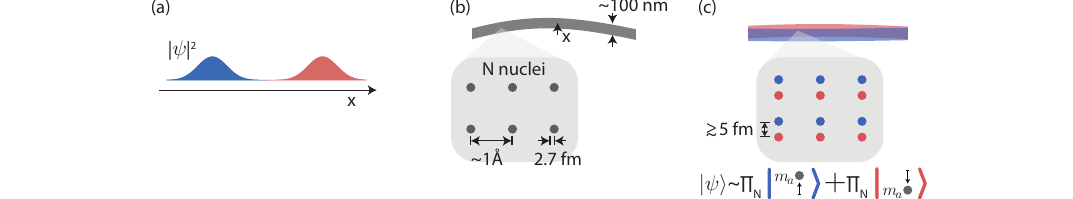}
\caption{\textbf{Implementation with mechanical oscillators.}
\textbf{(a)} The probability distribution $|\psi|^2$  of a cat state plotted in the position ($x$) basis.
The distribution features two maxima, and correspond approximatively to the gedanken experiment of Fig.~\ref{fig:time_dilation}.
For a physical system such as a membrane constituted of N nuclei \textbf{(b)}, these states can be viewed as a quantum superposition where the nuclei are simultaneously displaced up and down \textbf{(c)}.
}
\label{fig:superposition}
\end{figure*}

Although we previously gave the example of a statically displaced ball, we focus here on the common approach of realizing this experiment with a mechanical oscillator~\cite{marshall2003towards}.
This mechanical oscillator is given a total mass $m$, and an frequency $f_m = \omega_m/2\pi$.
The quantum state of this oscillator which will be most appropriate for measuring GR effects cannot be exactly predicted.
All we can say is that we are looking to prepare a pure quantum state with a large uncertainty in position such that we can approximate the GR time-scale $t_\text{GR}$ by Eq.~(\ref{eq:t_GRP}).
Cat states (Fig.~\ref{fig:superposition}(a)) are quantum states which most closely resemble the gedanken experiment of Fig.~\ref{fig:time_dilation}, and will be the states we consider.
We note however that any state featuring a large uncertainty in position would suffice, notably Fock states of motion, for which very similar formulas to the ones below can be derived.

Cat states are the closest analogue to the ball displaced in two directions.
A cat state is a superposition of coherent states oscillating out of phase such that when one part of the superposition is moving in one direction, the other part is moving in the other.
At maximum displacement (shown in Fig.~\ref{fig:superposition}(a)), the center of mass is localized in two positions separated by a distance $\Delta x = 2\sqrt{n}x_\text{zpf}$
where $n$ is the size of the cat given in number of phonons, and the zero-point fluctuations in position are given by 
\begin{equation}
    x_\text{zpf} = \sqrt{\frac{\hbar}{2\omega_m m}}\ .
    \label{eq:x_zpf}
\end{equation}
If we assume that coherence is limited by interaction with a thermal bath at temperature $T$, the time-scale on which a cat state will decohere to a mixture of two Gaussian states is given by~\cite{asjad2014reservoir} 
\begin{equation}
    t_\text{coh} = \frac{1}{2(2n_\text{th}+1)n\gamma_m}\ ,
    \label{eq:tcoh_cat_state}
\end{equation}
where $\gamma_m$ is the damping rate of the mechanical oscillator
and $n_\text{th} = 1/[\exp (\hbar\omega_m/k_BT)-1]$ is the average number of phonons occupying the oscillator after thermalization with the environment.
This coherence time is to be compared with the GR time-scale.
We have made the choice in this study to focus on the Penrose time-scale $t_\text{GR,P}$ of Eq.~(\ref{eq:t_G_penrose_full}).
When calculating this time for mechanical oscillators, it is however unclear what form the mass distribution $\rho$ should take.
In other words, it is unclear what the characteristic size $R$ of the oscillator should be, when making a parallel with the gedanken experiment of Fig.~\ref{fig:time_dilation}.
Note that the condition $\Delta x \gtrsim 2R$ is required to obtain a small and realistically attainable GR time-scale.
However, if we consider the oscillator to be a membrane, and use its thickness as a measure of $R$, then the quantum uncertainty in position $\Delta x$ will always be much smaller than $R$.
The membrane will be at least tens of nanometers thick, orders of magnitude larger than typical femtometer zero-point fluctuations.
Looking at the membrane at the atomic scale instead, the mass is concentrated in the nuclei, which may have a size much smaller than $\Delta x$, as illustrated in Fig.~\ref{fig:superposition}(b,c).
Using the nucleus radius as $R$ then yields a much shorter GR time-scale.
This approach can be justified by the common argument~\cite{kleckner2008creating} that the dominant mass of the atom is located in the nucleus.
See Appendix \ref{sec:appendix_R} for discussions on this choice of $R$, and on a recent experiment~\cite{donadi2020underground} placing a lower bound on $R$, for a given Di\'osi-Penrose collapse model.
In this approach, we also make the assumption that the thermal and quantum noise of all the other mechanical modes of the membrane are uncorrelated with the mode in which the superposition is constructed, which is justified by the extremely weak mechanical anharmonicity of mechanical oscillators at the energy scale of a few phonons.
For example, thermal excitation of another mode will randomly displace the nuclei in different directions, but in precisely the same way for the two parts of each nucleus superposition, therefore not affecting the above calculations of $t_\text{GR}$.

We will add the difference in gravitational self-energy of each nucleus to compute the GR time-scale.
We will thus require that the spatial extent of the superposition $\Delta x$ is larger than twice the nucleus radius
\begin{equation}
    a = A^{\frac{1}{3}}R_0\simeq 2.7\ \text{fm}\ ,
\end{equation}
where A is the mass number of the atom, $R_0 = 0.9$ fm~\cite{mohr2000codata} and the approximate equality holds for the mass and radii of common isotopes of silicon ($A=28$) and aluminum ($A=27$), typical materials for the fabrication of the mechanical oscillators we will be considering.
If $\Delta x\gtrsim2a$, the difference in gravitational self-energy following Eq.~(\ref{eq:delta_E_calculated}) is
\begin{equation}
    \Delta E \simeq \frac{48\pi Gm_a^2}{5a}\ .
\end{equation}
We have assumed that the nucleus is a ball of radius $a$ and mass $m_a = A m_u$ with $m_u\simeq1.7\times 10^{-27}$ kg.
Multiplying this energy difference by the number of nuclei in the mechanical oscillator,
we obtain the GR time-scale of the entire oscillator
\begin{equation}
    t_\text{GR} = \frac{1}{m}\frac{5 \hbar a}{48\pi Gm_a}\simeq \frac{1}{m}\ 3\times 10^{-15}\text{kg.s}\ ,
    \label{eq:gravitational_time-scale}
\end{equation}
%
where the approximate equality holds for silicon or aluminum.
In addition to $t_\text{coh}\gg t_\text{GR}$, we thus also require the spatial extent of the superposition $\Delta x$ to exceed the diameter of the nucleus $2a$.

\section{Candidate mechanical systems to detect GR effects}

\begin{figure*}[]
\centering
\includegraphics[width=0.9\textwidth]{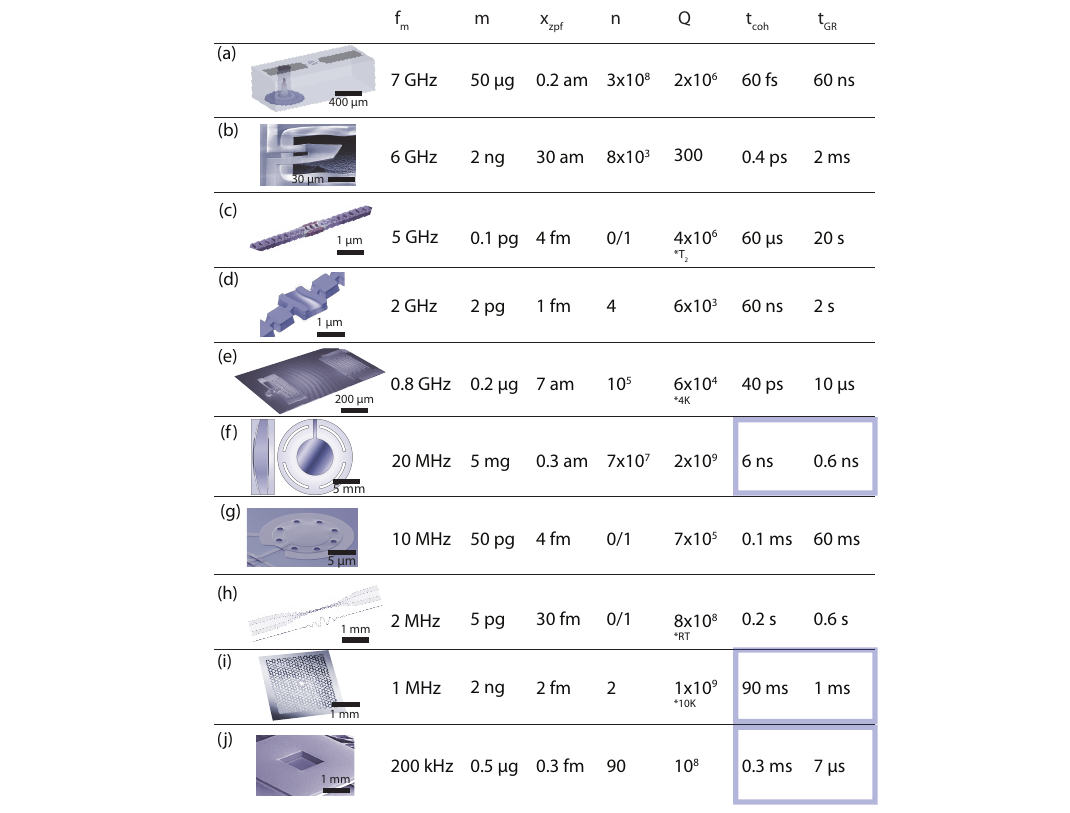}
\caption{
\textbf{Comparison of mechanical oscillators for the observation of GR effects.}
Systems where the coherence time $t_\text{coh}$ exceeds the GR time $t_\text{GR}$ are most favorable and highlighted in blue.
Note that $t_\text{coh}$ is a function of the quality factor $Q$, but also of the size of the cat state (given in number of phonons $n$), and the effect of the thermal environment.
The size of the state $n$ ensures that the superposition extends on a distance $\Delta x$ larger than the size of an atomic nucleus $2a$, where the mass is concentrated.
We write 0/1 for cases where small cat or Fock states ($n\sim1$), or even the ground-state ($n=0$) is sufficient to obtain $\Delta x\gtrsim 2a$.
In these cases, the coherence time is calculated for a cat state of size $n=1$.
For image credits and references for the quoted values, see Appendix ~\ref{sec:appendix_fig3}.
}
\label{fig:candidate_oscillators}
\end{figure*}

In this section we consider the suitability of different mechanical oscillators to explore GR effects in massive quantum superpositions.
We will solely focus on mechanical oscillators which have successfully been coupled to superconducting circuits: acoustic oscillators, beams and membranes or drums.
In order to assess each oscillator, we proceed as follows.
We first source the mechanical frequency $f_m$, mass $m$ and highest measured quality factor $Q = \omega_m/\gamma_m$ from literature (details in Appendix~\ref{sec:appendix_fig3}).
From $f_m$ and $m$, we compute the zero-point fluctuations of motion $x_\text{zpf}$ using Eq.~(\ref{eq:x_zpf}).
From $x_\text{zpf}$, we determine the size of the cat state, in number of phonons $n$, to implement the gedanken experiment schematically shown in Fig.~\ref{fig:superposition}. 
The state size $n$ should be large enough to lead to an uncertainty in position $\Delta x$ larger than the diameter of the atomic nucleus $2a$.
We then use $Q$ and $n$ to compute the coherence time $t_\text{coh}$ of Eq.~(\ref{eq:tcoh_cat_state}), assuming the oscillator is thermalized at $10$ mK.
Note that the higher $n$ is required, the shorter $t_\text{coh}$.
We compare $t_\text{coh}$ to the GR time $t_\text{GR}$ of Eq.~(\ref{eq:gravitational_time-scale}).
Each quantity corresponding to a step in this process is displayed in Fig.~\ref{fig:candidate_oscillators}.
We will now further discuss the results obtained for each type of oscillator.

Acoustic resonators, including phononic crystal defects~\cite{arrangoiz2019resolving} (Fig.~\ref{fig:candidate_oscillators}(d)), surface~\cite{satzinger2018quantum} (Fig.~\ref{fig:candidate_oscillators}(e)) and bulk~\cite{o2010quantum,chu2018creation} (Fig.~\ref{fig:candidate_oscillators}(a,b)) acoustic waves or optomechanical crystals~\cite{forsch2019microwave} (Fig.~\ref{fig:candidate_oscillators}(c)), are typically coupled to superconducting circuits through piezoelectricity.
The above-cited systems have frequencies in the GHz range, such that their coherence times are no longer limited by thermal effects when cooled down to $10$ mK in a dilution refrigerator.
By combining the conditions $(\Delta x)^2 \gtrsim (2a)^2$ and $t_\text{coh}\gg t_\text{GR}$, with $n_\text{th} = 0$, we find
\begin{equation}
     f_m\ 9\times 10^5 \text{ s}.\text{kg}^{-1}\lesssim \frac{n}{m}\ll  \frac{Q}{f_m} 3\times 10^{13}\text{ Hz}.\text{kg}^{-1}\ .
     \label{eq:gravity_condition_numerical_no_thermal}
\end{equation}
%
By plugging in $f_m = 1$ GHz into these relations, we find that the quality factor should be $Q\gg 10^{11}$ to enter a regime where GR effects may become relevant, which is far from the current state of the art.

A more favorable acoustic system can be constructed by using the low-frequency modes of a $\sim$ 10 millimeter sized bulk-acoustic resonator~\cite{goryachev2014extremely,carvalho2019} (Fig.~\ref{fig:candidate_oscillators}(f)).
Both the effective mass and the quality factor of the acoustic resonance are then increased, at the expense of lowering the resonance frequency to $\sim$ 10 MHz where the mechanical mode will be thermally populated.
The zero-point fluctuations of motion are still relatively small (less than an attometer), such that states with more than a million phonons are required for the quantum uncertainty in position to exceed the size of the nucleus.
Such states will likely be challenging to construct.
However, if this is achieved, the large quality factor $\sim\ 10^9$ leads to a thermal coherence time of such states an order of magnitude longer than the GR time-scale $t_\text{GR}\sim0.6$ ns.
Another class of mechanical oscillators are beams, which can be coupled capacitively~\cite{regal2008measuring,barzanjeh2019stationary} or through superconducting quantum interference~\cite{rodrigues2019coupling} to superconducting circuits.
State of the art beams are made from non-uniform phononic crystals~\cite{ghadimi2018elastic} (Fig.~\ref{fig:candidate_oscillators}(h)), which are representative of typical beams in terms of size, but feature an extremely high quality factor $Q = 8\times 10^8$ (at room temperature).
Their zero-point fluctuations are large enough for $\Delta x \gtrsim 2a$ even in the ground-state.
However, their mass is too low to have a sufficiently short GR time-scale, and $t_\text{GR}\ll t_\text{coh}$ would require $Q \gtrsim 10^{10}$.
The last type of oscillators are membranes. 
High frequency membranes (also known as drums) can be made by suspending thin films of superconductor, usually aluminum (Fig.~\ref{fig:candidate_oscillators}(g)).
These typically resonate at $f_m\sim 10$ MHz, and have relatively small mass $m\sim 50$ pg~\cite{Teufel2011} and quality factors $Q \sim 7\times 10^5$ ~\cite{clark2017sideband}.
As a consequence, their GR time-scale $t_\text{GR}\sim 60$ ms remains inaccessible considering their typical thermal coherence time $t_\text{coh}\sim 100\ \mu$s.
Due to their large mass ($m=500$~ng~\cite{yuan2015large}), 1 mm sized high-stress SiNx membranes (Fig.~\ref{fig:candidate_oscillators}(j)) have short GR time-scales of $t_\text{GR}\sim 7\ \mu$s. 
Combined with their low-frequency ($200$ kHz), their zero-point fluctuations $x_\text{zpf}\sim 0.3$ fm are only an order of magnitude smaller than the size of a Si nucleus $2.7$ fm.
With large measured quality factors ($10^8$), $t_\text{GR}$ is much shorter than the thermal coherence time $t_\text{coh}\sim 0.3$ ms of Fock or cat states of amplitude $n\sim100$.
Reducing the size and mass of these membranes will increase the GR time-scale, but through the use of phononic shields~\cite{yu2014phononic}, soft clamping and dissipation dilution~\cite{tsaturyan2017ultracoherent}, this can be compensated by an increase in quality factor.
Smaller mass also translates to larger zero-point fluctuations, diminishing the size of cat state which is necessary.
An order of magnitude higher frequency resonators,  $f_m = 1$ MHz, with two orders of magnitude smaller effective mass $m=2$~ng were demonstrated to have $10^9$ quality factors at 10 Kelvin~\cite{chen2020entanglement} (Fig.~\ref{fig:candidate_oscillators}(i)). 
Their GR time-scale $t_\text{GR}\sim 1$ ms, is then much shorter than the thermal coherence time $t_\text{coh}\sim 90$ ms of low amplitude Fock or cat states.
Such states are sufficient here as the zero-point fluctuations $x_\text{zpf}\sim 2$ fm are on the order of the size of a Si nucleus $2.7$ fm.

Overall, we have found that low-frequency oscillators tend to be most adapted to detecting GR effects.
One way to understand why this emerges from our analysis is to combine the conditions $(\Delta x)^2 \gtrsim (2a)^2$ and $t_\text{coh}\gg t_\text{GR}$ into one relation, in the assumption $k_B T \gg \hbar \omega_m$
\begin{equation}
     f_m\ 9\times 10^5 \text{s.kg}^{-1}\lesssim \frac{n+\tfrac{1}{4}}{m}\ll  Q\  6 \times 10^{4}\text{ kg}^{-1}\ ,
     \label{eq:gravity_condition_numerical}
\end{equation}
where the extra ``$+\tfrac{1}{4}$'' captures the coherence time $1/(\gamma_m n_\text{th})$ and position uncertainty $\Delta x = x_\text{zpf}$ of the ground state for the case $n\rightarrow 0$ occurring for example in Fig.~\ref{fig:candidate_oscillators}(h).
Looking at the left-most and right-most terms, we see that an increase in frequency requires an increase in quality factor.
Conversely, low-frequency oscillators will impose less stringent requirements on the quality factor, and are more favorable candidates to measuring GR effects.
Quantitatively, we see that $100$ kHz requires at least $Q\gg 10^6$, $1$ MHz requires $Q\gg 10^7$, etc...

\section{Conclusion}
In conclusion, we have laid out a method to evaluate the potential of mechanical oscillators to measure GR effects in massive superpositions.
In the case of oscillators which can be coupled to superconducting circuits, we have determined that low-frequency silicon nitride membranes, manipulated into small cat-states or Fock-states, or low-frequency acoustical resonators, but manipulated into quantum states with over a million phonons, are most appropriate.
The challenge in performing such experiments currently lies in gaining quantum control over these objects, for example by extending the techniques already developed~\cite{o2010quantum,satzinger2018quantum,chu2018creation,arrangoiz2019resolving,ma2020nonclassical,ockeloen2018stabilized,reed2017faithful,gely2021phonon}, to lower mechanical frequencies.
Other systems could be considered for such experiments, notably levitated nano- or micro- spheres. 
Recently, a milestone experiment towards constructing superposition states in these systems was performed: the motion of an optically levitated nano-sphere was successfully cooled to the quantum ground state~\cite{Delic}. 
However, a calculation of the GR time-scale of the cooled 200 nm diameter silica sphere, using Eq.~(\ref{eq:gravitational_time-scale}), shows that $t_\text{GR}\sim300$ s, greatly exceeding its estimated maximum coherence time of $7.6\ \mu$s.
Theoretical proposals, based on magnetic rather than optical levitation for example~\cite{PhysRevLett.109.147205}, predict that similar level of control could be brought to larger, micro-spheres, with shorter GR time-scales $t_\text{GR}\sim$ms, and near-infinite (hundreds of hours) coherence times.
Many assumptions have been taken to reach the conclusions of this article.
We have assumed that $E=mc^2$ is the correct energy-scale in predicting a GR time-scale $t_\text{GR}$.
We have assumed that it is sufficient to create a superposition with an uncertainty in position larger than the diameter of the nucleus, on the grounds that the mass is concentrated there.
We have assumed that the coupling to the thermal environment would govern the coherence time of a quantum state of motion.
And even if one could sustain a quantum state on a time-scale $t_\text{GR}$, we lack a theory predicting the measurable consequences of the time-dilation superposition.
However, we hope that the ideas laid out in this article will encourage experimentalists to try and gain quantum control over lower-frequency massive oscillators; as even sustaining a quantum superposition for a time $t_\text{GR}$ without measurable consequences could advance our understanding of how quantum mechanics and general relativity interplay by disproving some of the ideas discussed here.
\vspace{5pt}

\textbf{Acknowledgments}
The authors acknowledge funding from the European Union’s Horizon 2020 research and innovation programme under grant agreement 681476 - QOM3D.
\vspace{5pt}

\textbf{Data Availability}
Data sharing is not applicable to this article as no new data were created or analyzed in this study.
The code used in generating the table of Fig.~\ref{fig:candidate_oscillators} is available in Zenodo with the DOI identifier 10.5281/zenodo.4625431.

\begin{appendix}
\section{On the choice of $R$ and the recent experiment ``Underground test of gravity-related wave function collapse''}
\label{sec:appendix_R}
Note that the characteristic size $R$ of the object in superposition is critical in determining both the gravitational time-scale $t_\text{GR}\propto R$, as well as the minimum extent in space a quantum superposition should have $\Delta x \gtrsim 2R$.
Unfortunately, our choice of $R$ being equal to the size of the nucleus is at best an educated guess.
An alternative would have been to take the extent of the ground-state wavepacket $x_\text{zpf}$~\cite{kleckner2008creating}. 
Since $x_\text{zpf}$ is typically of the same order or smaller than the nucleus radius (see Fig~\ref{fig:candidate_oscillators}), this choice would have given more favorable numbers for the higher frequency oscillators.
Recently, an experiment was carried out in order to place a lower bound on $R$~\cite{donadi2020underground}.
In the assumption that, on the GR time-scale, we have Poissonian collapse of the wave function, gravitational collapse should lead to radiation.
This radiation was not measured, and the precision of the measurement leads to the bound $R>5\times 10^{-11}$.
This bound is four orders of magnitude above the size of the nucleus used in our discussion.
And repeating the calculations of Fig.~\ref{fig:candidate_oscillators} with this lower bound reveals that none of these mechanical oscillators are now suitable to explore \textit{the particular gravitational collapse model} considered in the experiment of Ref.~\cite{donadi2020underground}.
This does not exclude the appearance of different consequences of general relativity in massive superpositions on the time-scale $t_\text{GR}$ with $R$ equal to the nuclear radius.

\section{Construction of Fig.~3}
\label{sec:appendix_fig3}
Illustrations reprinted with permission from: (a): Y. Chu, P. Kharel, W. H. Renninger, L. D. Burkhart, L. Frunzio, P. T. Rakich, and R. J. Schoelkopf, Science 358, 199–202 (2017). Copyright 2017 AAAS; (b): A. D. O’Connell, M. Hofheinz, M. Ansmann, R. C. Bialczak, M. Lenander, E. Lucero, M. Neeley, D. Sank, H. Wang, M. Weides, et al., Nature 464, 697–703 (2010). Copyright 2010 Springer Nature; (c): R. Riedinger, S. Hong, R. A. Norte, J. A. Slater, J. Shang, A. G. Krause, V. Anant, M. Aspelmeyer, and S. Gröblacher, Nature 530, 313–316 (2016). Copyright 2016 Springer Nature; (d): P. Arrangoiz-Arriola, E. A. Wollack, Z. Wang, M. Pechal, W. Jiang, T. P. McKenna, J. D. Witmer, R. Van Laer, and A. H. Safavi-Naeini, Nature 571, 537 – 540 (2019). Copyright 2019 Springer Nature; (e): P. Delsing, A. N. Cleland, M. J. Schuetz, J. Knörzer, G. Giedke, J. I. Cirac, K. Srinivasan, M. Wu, K. C. Balram, C. Bäuerle, et al., J. Phys. D Appl. Phys. 52, 353001 (2019), under a Creative Commons Attribution 3.0 license; (f): M. Goryachev, D. L. Creedon, E. N. Ivanov, S. Galliou, R. Bourquin, and M. E. Tobar, Applied Physics Letters 100, 243504 (2012). Copyright 2012 AIP Publishing; (g): J. B. Clark, F. Lecocq, R. W. Simmonds, J. Aumentado, and J. D. Teufel, Nature 541, 191–195 (2017); (h): A. H. Ghadimi, S. A. Fedorov, N. J. Engelsen, M. J. Bereyhi, R. Schilling, D. J. Wilson, and T. J. Kippenberg, Science 360, 764–768 (2018). Copyright 2018 AAAS; (i): https://www.nbi.ku.dk/english/news/news17/holey-pattern-boosts-coherence-of-nanomechanical-membrane-vibrations/; (j): M. Yuan, V. Singh, Y. M. Blanter, and G. A. Steele, Nat. Commun. 6, 8491 (2015), under a Creative Commons Attribution 4.0 International License.

Numbers are quoted from the above references except:
(a) Mass obtained from the volume under the piezoelectric disk. Quality factor quoted from  Ref.~\cite{chu2018creation}.
(b,d) Mechanically compliant volume and effective mass extracted from the thickness of the deposited materials, and an estimate of the cantilever/defect area of motion. 
(c) Numbers quoted from Ref.~\cite{maccabe2019phononic}. For this system, the decoherence rate was measured, in addition to the decay rate. The former is the relevant number quoted here.
(e) Effective mass extracted from the zero-point fluctuations and frequency quoted in Ref.~\cite{noguchi2017qubit} . Quality factor quoted from Ref.~\cite{shao2019phononic}.
(g) Frequency and mass quoted from Ref.~\cite{Teufel2011}.
(h) System measured at room temperature.
(i) Numbers quoted from Ref.~\cite{chen2020entanglement}, measured at 10 Kelvin.
(j) Numbers quoted from Ref.~\cite{yuan2015silicon}.
For mechanical oscillators where the size of state is $n=0/1$, we computed the coherence time of a cat state of size 1.
For a thermally populated oscillator, this is close the coherence time of the ground-state $1/(n_\text{th}\gamma)$, which in these cases presents a large enough position uncertainty to explore GR effects.
\end{appendix}

\bibliography{library}
    
\end{document}